# Coherent Aharonov-Bohm oscillations in type-II (ZnMn)Te/ZnSe quantum dots

I. R. Sellers[1*], V. R. Whiteside[1], A. O. Govorov[2], W. C. Fan[3], W-C. Chou[3], I. Khan[1], A. Petrou[1] and B. D. McCombe[1].

[1]*Dept. of Physics, Fronzcak Hall, University at Buffalo SUNY, Buffalo, NY 14260, USA*

[2]*Dept. of Physics & Astronomy, Ohio University, Athens OH 45701, USA.*

[3]*Department of Electrophysics, National Chiao Tung University, Taiwan, Republic of China.*

**March 2008**

The magneto-photoluminescence of type-II (ZnMn)Te quantum dots is presented. As a result of the type-II band alignment Aharonov-Bohm (AB) oscillations in the photoluminescence intensity are evident, confirming previous predictions for the suitability of this geometry to control the optical Aharonov-Bohm effect. Moreover, the system demonstrates an interesting interplay between the AB effect and the spin polarization in diluted magnetic semiconductor quantum dots. The intensity of the AB oscillations increases with both magnetic field and the degree of optical polarization, indicating the suppression of spin fluctuations improves the coherence of the system.

**[*]isellers@buffalo.edu**

The Aharonov-Bohm (AB) effect describes a phase shift induced upon a charged particle as it moves in a closed trajectory in the presence of a magnetic field [1]. Despite the fact that the AB effect is a property of charged particles it has been shown that for neutral excitons in semiconductor nano-rings, a non-zero electric dipole moment exists [2, 3], which is adequate to create AB oscillations. Such behavior results from the inherent differences in confinement potential and effective mass of the particles comprising the exciton, the electron and hole, and leads to different trajectories for each of the charge carriers, resulting in a measurable AB effect. Type-II quantum dots (QDs) have been predicted to be particularly amenable to exhibiting such AB effects because of the enhanced polarization of the electron and hole due to the spatial separation of the carriers in such systems [3, 4].

In this work we present experimental evidence of AB oscillations in the magneto-photoluminescence (MPL) of type-II (ZnMn)Te/ZnSe QDs. In this system the hole is strongly confined within the (ZnMn)Te QD while the electron resides in the ZnSe matrix, confined only through coulomb attraction to the hole. Although the AB effect has been observed optically for charged excitons in In(Ga)As QDs [5] and neutral excitons in both Type-II GaAs/InP [6] and Zn(SeTe) QDs [7, 8], and also in capacitance [10] and magnetization measurements [11] in InAs quantum rings, this is the first time such effects have been observed in a diluted magnetic semiconductor (DMS) system. In DMS systems the application of a magnetic field strongly aligns the Mn spins, thus polarizing the emission through the carrier-Mn exchange [12]. At saturation polarization the carrier spins will also be preferentially aligned, significantly reducing the spin disorder in the system [9, 10].

The samples studied were grown by MBE on (001) GaAs substrates. The GaAs buffer layer was planarized at 580ºC before the temperature was reduced to 300ºC to deposit a ZnSe buffer. The active layers were grown by migration enhanced epitaxy initiated by the growth of several mono-layers of ZnSe followed by the deposition of the (ZnMn)Te QDs. In the sample described here, five 2.6 ML (ZnMn)Te QD layers were grown, separated by narrow (5 nm) ZnSe spacer layers. The QD density of each layer was estimated from atomic force microscopy images to be of the order of $10^9$ – $10^{10}cm^{-2}$. During growth the Mn-composition in the QDs was estimated to be ~11%. The growth optimization and material characterization of this sample is described fully elsewhere [11].

The sample was mounted in the Faraday geometry and excited with the 488 nm line of an $Ar^+$ laser at 4.2 K. Figure 1 shows the photoluminescence (PL) at 4.2K. The QD emission peaks at ~1.92 eV, which is much lower in energy than the peak emission from bulk $Zn_{0.89}Mn_{0.11}Te$ at low temperature (~2.42 eV) [12, 14]. This large redshift of the energy gap observed for the (ZnMn)Te/ZnSe QDs is a direct result of the type-II band alignment and confinement potential of the dots.

The incorporation of Mn into the QDs forms the paramagnetic alloy (ZnMn)Te due to the large exchange interaction that exists between Mn and the charge carriers in this material [9]. These properties result in a large optical polarization of the PL with applied magnetic field. This behavior is illustrated as an inset (a) to Fig. 1, which shows the integrated intensity of the $S^+$ and $S^-$ emission. As the magnetic field increases the electron ($m_s$ = ±1/2) and heavy hole states ($m_j$ = ±3/2) split, creating a situation where the carriers preferentially occupy their lowest energy states ($m_s$ = -1/2, $m_j$ =3/2). The emitted PL then directly represents the total carrier spin orientation due to the selection rules of the transition [10]. Here, since the dominant transition is associated with an exciton formed between (spin down) electrons ($m_s$ = -1/2) and (spin up) holes ($m_J$ = +3/2), with total angular momentum projection equal to +1 [10], the emission of photons is dominated by left-circularly polarized ($S^+$) light.

Figure 1(b) shows the optical polarization, $P = (I_+ - I_-)/(I_+ + I_-)$, where $I^+(I^-)$ represents the intensity of the $S^+$ ($S^-$) luminescence, respectively, with increasing magnetic field. The optical polarization degree is greater than 95% at magnetic fields above 4T. At saturation, the carrier spins align preferentially due to exchange interaction with the Mn spins and this significantly reduces the spin disorder in the system, the significance of which is discussed below.

Also shown as inset (c) to Fig. 1 is a schematic representation of the sample structure, which illustrates the self-assembly of the QD layers into columns. This structure was confirmed by cross-sectional transmission electron microscopy [11], and occurs via the strain interaction between the QD layers due to the relatively narrow ZnSe spacer regions that separate them. The columnar geometry of the type-II QDs has previously been shown to be particularly amenable to observe the optical AB effect, since not only is the electron *bound* to the *confined* hole through coulomb attraction, but the electron also rests preferentially in-plane, thus defining a ring-like geometry. This is shown schematically as an inset to Fig. 2(b). In such structures the electron sits laterally besides the hole, confined by a combination of the coulombic attraction and the narrower ZnSe spacer layer between the QD planes.

The main panel in Figure 2(a) shows the magnetic field dependence of the PL intensity for an ensemble of (ZnMn)Te/ZnSe QD columns. The overall PL intensity increases with magnetic field consistent with the "squeezing" of the electron wavefunction at the QD interface, which increases the electron-hole wavefunction overlap and thus the oscillator strength. The increase in PL intensity is illustrated further by the upper inset (left) to Fig. 2(a). This shows the PL spectra with increasing magnetic field, which are integrated to establish the total spontaneous emission. In the main panel of Fig. 2(a) it is also evident that periodic oscillations are visible superimposed upon the rising intensity background of the emission. These oscillations are clear evidence of a coherent AB effect between the constituent particles of the exciton; the electron and hole, and are consistent with similar behavior in Zn(SeTe) QDs [7, 8].

The specific origin of oscillatory behavior in the PL intensity is related to the cylindrical symmetry of the QD column and the optical AB effect. The radial spatial separation in the type-II QD creates a rotating dipole with the two charged particles of the exciton orbiting over different areas. This behavior is described qualitatively by a simple model of a rotating dipole with a magnetic field normal to the plane;

$$E_{exc} = E_g + \frac{\hbar^2}{2MR_0^2}\left(L + \frac{\Delta\Phi}{\Phi_0}\right)^2, \qquad (1)$$

where $L$ is the angular momentum quantum number, $E_g$ is the confined hole- to-electron ground state energy, $\Phi_0 = hc/|e|$ is the magnetic flux quantum, $R_0 = (R_e + R_h)/2$, and $M = (m_e R_e^2 + m_h R_h^2)/R_0^2$; $R_h$ and $R_e$ are the averaged radii of orbits of the hole and electron respectively, and $m_h$ and $m_e$ are their masses. Here, since the hole is strongly localized, $R_h = 0$. The quantity $\Delta\Phi = p(R_e^2 - R_h^2)B = 2p\Delta R \cdot R_0 B$ is the magnetic flux through the area between electron and hole trajectories. The inset to Fig. 3 shows a simple schematic of the orbital motion of particles inside such a localized exciton.

With increasing magnetic field the $L$ value characterizing the ground state angular momentum projection changes from $L = 0$ to a non-zero value of $L$, due to the cylindrical symmetry of the system. Such behavior should have significant impact upon the PL intensity since the selection rules for the optical transition are modified, and in the simplest case, with increasing magnetic field when $L$ is non-zero, the intensity should be strongly suppressed [3]. Although this is not totally consistent with the experimental data (Fig 2); since the system is not perfectly cylindrical, the selection rules will be relaxed, and consequently excitons with non-zero $L$ will have finite optical oscillator strength [4], as observed.

In Fig. 2(a) the AB oscillation period (*d*B) at B > 5 T can be estimated as 1.1 T. The resultant effective radius ( $R_e = \sqrt{\Phi_0 / pdB}$ ) of the electron is determined to be ~28 nm, which is larger than the 20 nm lateral extent of the QDs as determined by structural analysis [11]. This difference is consistent with the picture of the electron orbiting the QD perimeter bound to the hole [2, 4]. Importantly, the estimated AB oscillations in the energy of the lowest state determined using Equation (1) of the exciton is ~ 0.04 meV, due to the relatively large lateral radius ($R_e$) of the electron orbit (28 nm). As a result, we do not expect to observe these weak energy oscillations experimentally in the peak PL energy due to the strong inhomogeneous broadening (~80 meV) of the excitonic peak. This is indeed shown to be the case in Figure 3, where the position of the PL peak does not exhibit any visible oscillations, but rather a slight reduction in the transition energy with magnetic field. The origin of this behavior is related to the Zeeman contribution to the emission, the behavior of which is discussed in Ref [13].

Despite the absence of AB oscillations in the PL energy, the strength of the oscillations in the emission intensity increases with magnetic field. This can be seen more clearly in the lower (right) inset to Figure 2(a), which shows the peak intensity of the oscillations with the broad featureless PL background subtracted. In this figure three oscillations are clearly evident with peaks at ~ 5, 7 and 9 T. This enhancement of the AB effect at higher magnetic fields appears to be strongly related to the spin polarization of the carriers in the DMS QDs (and thus to the magnetization of the dots) and can be correlated with the optical polarization (see inset (b) Fig. 1). At low B-fields the Mn spins are randomly orientated and fluctuating due to the finite temperature of the system (~4 K). This is consistent with the low paramagnetic to spin-glass phase transition temperature of dilute ZnMnTe [9, 12] and, specifically, the relatively low freezing temperature expected for the Mn composition (11%) of the QDs studied here (~0.4 K for bulk $Zn_{0.93}Mn_{0.07}Te$) [12].

Since the electron in an exciton and the Mn-spins are coupled through an exchange interaction, fluctuations of Mn spins create a fluctuating exchange potential for the electron orbiting a QD column. Such fluctuations may have a destructive effect upon the AB phase, which may not survive. The behavior of the spectra in Fig. 2(a) at B < 4 T is consistent with this idea. As the magnetic field aligns the Mn spins, magnetic fluctuations in the system are reduced and the AB oscillations increase in strength. This is also consistent with the increase in the optical polarization degree, with the largest AB oscillations observed at magnetic fields above polarization saturation (> 4 T).

The average magnetic moment of a single Mn-impurity and the strength of spin fluctuations can be written as

$$\overline{M}_z = \langle M_z \rangle = S \cdot g_{Mn} \cdot m_B \cdot B_{5/2}(x) \text{ and } \sqrt{dM^2} = \sqrt{\left\langle \left(M_z - \overline{M}_z\right)^2 \right\rangle} = \sqrt{k_B T \frac{\partial \overline{M}_z}{\partial B}}, \quad (2)$$

where $\langle M_z \rangle$ is the thermal average over the states of Mn spin with $S = 5/2$ ; $\boldsymbol{B}_{5/2}(\boldsymbol{x})$ is the Brillouin function, $x = g_{Mn} \cdot S \cdot m_B \cdot B / k_B T$ , and $B$ is the magnetic field. We take $T = 4.2K$ and $g_{Mn} = 2$ . The results of these calculations are shown in Fig. 2(b) and clearly illustrate the suppression of spin fluctuations for $\boldsymbol{B} > 4\boldsymbol{T}$ . Consequently, these predictions support the assumption that the increase of AB oscillations at higher $B$ is related to the reduction of the exchange potential fluctuations.

In summary we have presented clear evidence of the Aharonov-Bohm effect in the PL intensity of type-II (ZnMn)Te/ZnSe QDs. Simultaneously, we have observed an interesting interplay between the AB oscillations and the magnetization. The strength of the AB effect appears to be correlated with the degree of optical polarization of the system. We believe that with increasing magnetic field, and therefore, increased spin polarization, the AB oscillations become enhanced due to the suppression of spin fluctuations and related decoherence. The appearance of such oscillations in the columnar QDs presented here appears to confirm previous predictions of the suitability of this particular geometry [7, 8] for the creation and control of AB effects in semiconductor systems.

We acknowledge support from the Center for Spin Effects and Quantum Information in Nanostructures (UB).

**Figure 1.**

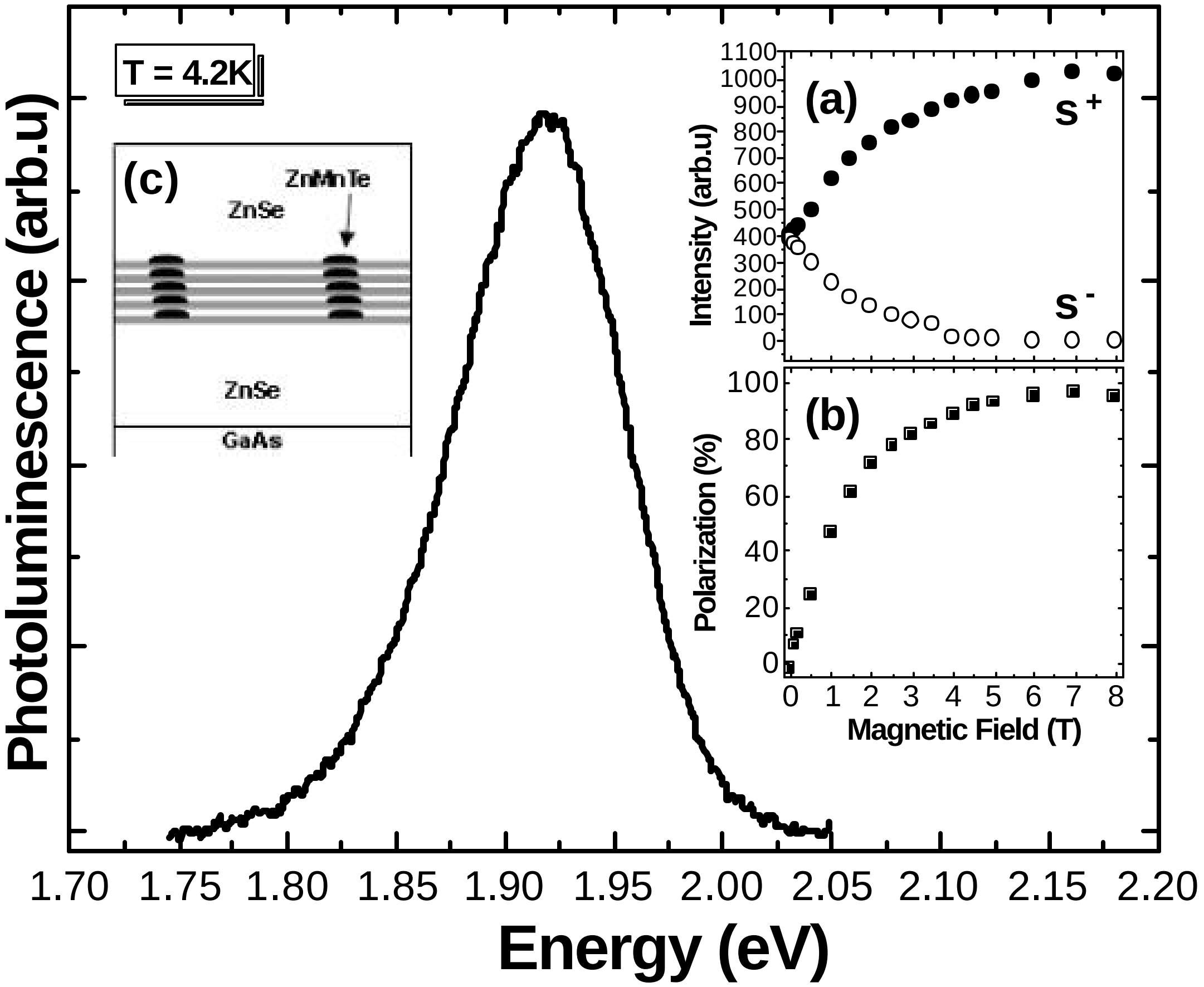

T = 4.2K
(c)
ZnMnTe
ZnSe
ZnSe
GaAs
(a)
s+
s-
Intensity (arb.u)
1100
1000
900
800
700
600
500
400
300
200
100
0
(b)
Polarization (%)
100
80
60
40
20
0
0 1 2 3 4 5 6 7 8
Magnetic Field (T)
Photoluminescence (arb.u)
1.70 1.75 1.80 1.85 1.90 1.95 2.00 2.05 2.10 2.15 2.20
Energy (eV)

**Figure 2.**

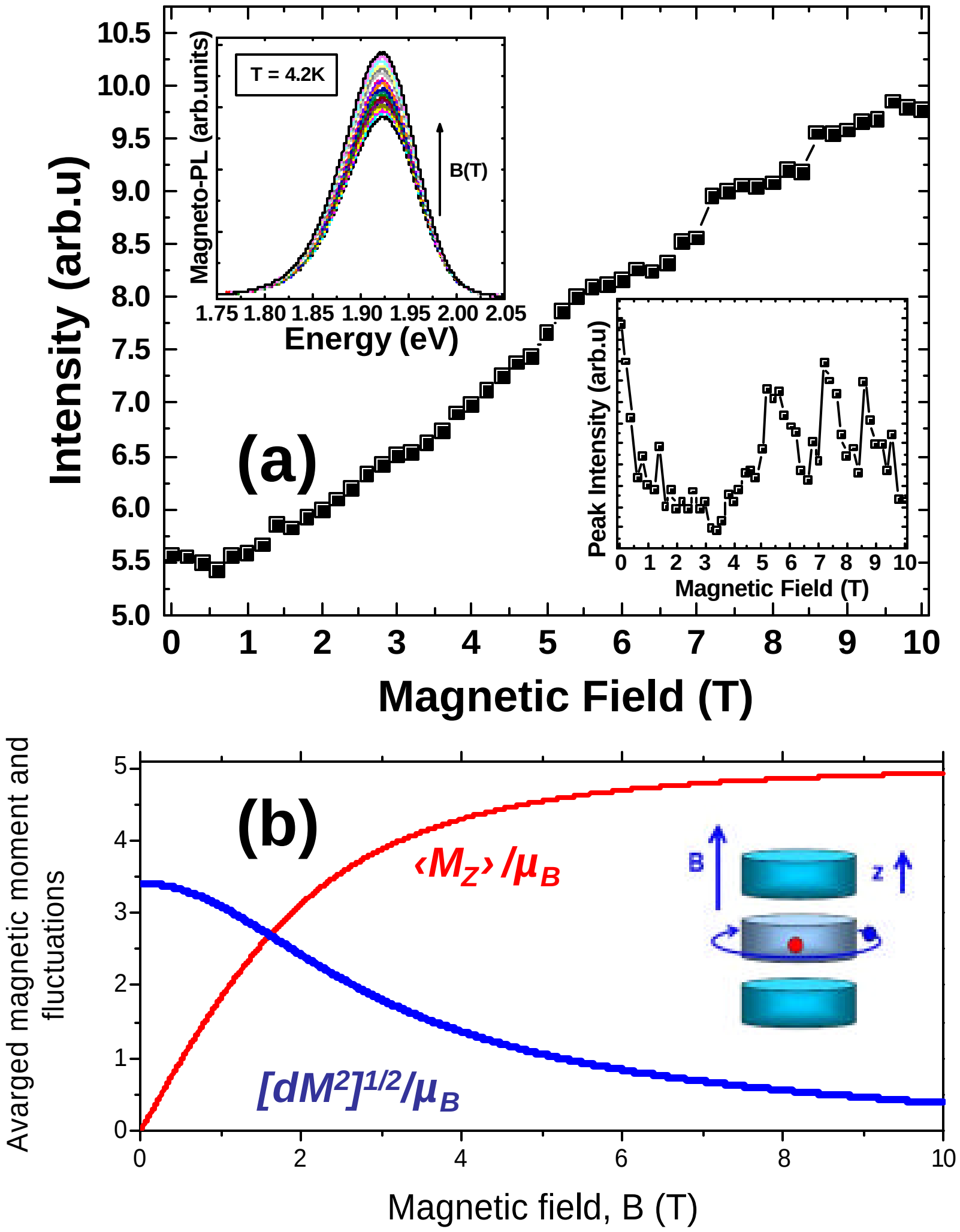

**Figure 3**

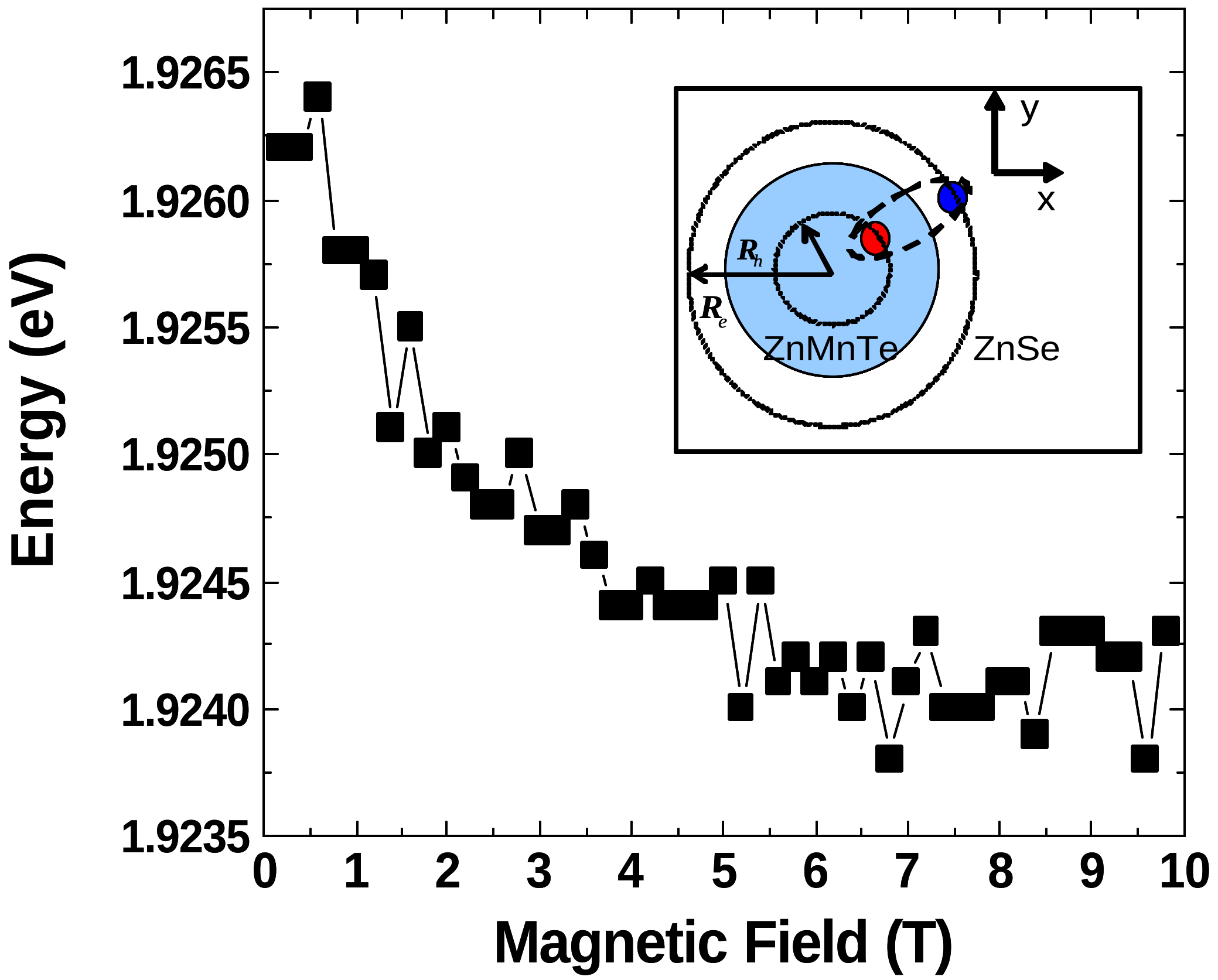

# Figure Captions

**Figure 1. PL of the ZnSe/(ZnMn)Te quantum dots at 4.2K. The inset (a) shows the intensity of the $S^+$ (full circles) and $S^-$ (open circles) photoluminescence component versus magnetic Field. The inset (b) shows the degree of optical polarization while inset (c) shows a schematic representation of the structure.**

**Figure 2. (a) Integrated PL intensity versus magnetic field for the ZnSe/(ZnMn)Te QDs at 4.2K. The upper inset (left) shows the magneto-photoluminescence spectra with increasing magnetic field. The lower inset (right) shows the peak intensity of the oscillations after removal of the featureless background. (b) Calculated average magnetic moment (red) and fluctuations of a single Mn-spin (blue) in the units of the Bohr magneton. The inset shows the columnar geometry of the stacked QDs.**

**Figure 3. PL energy versus magnetic field for the ZnSe/(ZnMn)Te QDs. The inset represents the simple model of a rotating dipole created by the electron and hole in this type-II columnar geometry.**